# Workload Scheduling on heterogeneous Mobile Edge Cloud in 5G networks to Minimize SLA Violation

Mostafa Hadadian Nejad Yousefi[1], Amirmasoud Ghiassi[1], Boshra Sadat Hashemi[1], Maziar Goudarzi[1]

## Abstract

Smart devices have become an indispensable part of our lives and gain increasing applicability in almost every area. Latency-aware applications such as Augmented Reality (AR), autonomous driving, and online gaming demand more resources such as network bandwidth and computational capabilities. Since the traditional mobile networks cannot fulfill the required bandwidth and latency, Mobile Edge Cloud (MEC) emerged to provide cloud computing capabilities in the proximity of users on 5G networks. In this paper, we consider a heterogeneous MEC network with numerous mobile users that send their tasks to MEC servers. Each task has a maximum acceptable response time. Non-uniform distribution of users makes some MEC servers hotspots that cannot take more. A solution is to relocate the tasks among MEC servers, called Workload Migration. We formulate this problem of task scheduling as a mixed-integer non-linear optimization problem to minimize the number of Service Level Agreement (SLA) violations. Since solving this optimization problem has high computational complexity, we introduce a greedy algorithm called MESA, Migration Enabled Scheduling Algorithm, which reaches a near-optimal solution quickly. Our experiments show that in the term of SLA violation, MESA is only 8% and 11% far from the optimal choice on the average and the worst-case, respectively. Moreover, the migration enabled solution can reduce SLA violations by about 30% compare to assigning tasks to MEC servers without migration.

## 1 Introduction

Smartphones have made their way to almost everybody life. They made their impact on all age groups. Mobile devices offer various services such as staying in touch with the world, full-house entertainment, and cheaper education. Mobile Internet enables consumers to access and share information on the go.

According to Statista report on January 2018, the global mobile population amounted to 3.7 billion unique users [1]. Besides, global mobile data traffic is about 19 EB/m (exabytes per month) and predicted to become 77 EB/m by 2020 [2]. Furthermore, part of this massive data traffic is because of latency-sensitive and high-performance application such as augmented reality (AR), autonomous driving [3], and online gaming. AR traffic will reach 4.02 EB/m by 2022, up from 0.33 EB/m in 2017 and the size of the market will grow from 6 billion U.S. dollars to 20 billion [4], [5]. The consumer data traffic in the online gaming segment is expected to rise from 1 to 15EB/m from 2017 to 2020 [6]. Table 1 illustrates the properties of applications. Table 1 shows the requirements for latency-sensitive applications. Accordingly, these types of applications need resources such as network bandwidth and computational capabilities. Traditional mobile networks cannot fulfill the required bandwidth and latency. Other than network latency, computation latency also plays a significant role in the total latency. These highly demanded computations drain the battery on mobile devices. Thus, using external and proximate resources make it possible to have a longer battery life for mobile users and higher QoE.

**Table 1** Properties of Latency-sensitive apps

| Application | Maximum deadline(ms) | Size of Input File (MB) |
|---|---|---|
| Augmented Reality[7], [8] | 75 | 1-7 |
| Online Gaming[7], [9] | 100 | 1-10 |
| Face Recognition[10] | 100 | 0.09-7.5 |
| Web Accelerate Browser[11] | 100-800 | 0.3-5 |
| Big Data analysis[12] | 200-900 | 0.1-1 |

Mobile Edge Cloud (MEC) is an abstraction that provides cloud computing capabilities in the proximity of users on the 5G network. MEC brings computation and storage resource to the edge of the network, enabling it to run the highly demanding applications at the user equipment while meeting strict delay requirements [13]. Figure 1 shows its architecture. Mobile users (MU) are scattered in regions and send their requests to nearest MEC servers that provide some services. MUs send their demands to the edge. The edge layer or the central data center can respond to the request. If a server receives a task and is not able to process it, it can redirect the task to either another server or the central data center. The process of transferring a task between MEC servers is called Workload Migration. Thus, using MEC can reduce the average response latency compared to either executing the workload on the mobile device or sending it to the central cloud.

[1] Department of Computer Engineering, Sharif University of Technology, Tehran, Iran

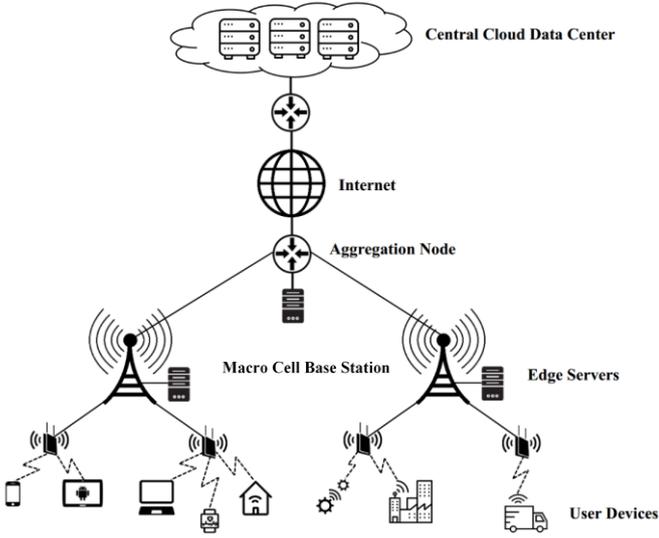

**Fig.1** Mobile Edge Cloud Conceptual Architecture

The MEC usage scenario of our interest is as follows: First, MU wants to send a task to the MEC server. Focusing on deciding whether the task should be offloaded to MEC server or not is called "Workload offloading" that is vastly studied [14]–[17]. The second step is to decide to offload the workload to which MEC server. To do this step, we can consider many factors such as mobility [18] and energy awareness management [19], and task scheduling among MEC servers [20]. These factors can be categorized as "Resource Management" in MEC servers.

Our work focuses on resource management and task scheduling to improve user satisfaction on heterogeneous 5G networks that use the MEC concept. In this work, we schedule tasks that are offloaded by MUs on MEC servers. We formulate this problem as a mixed-integer non-linear programming (MINLP). Then, according to the high computational complexity of solving MINLP problem to reach the optimal solution, we suggest a greedy algorithm called MESA, Migration Enabled Scheduling Algorithm. MESA provides near-optimal solutions, and The average difference between MESA and the optimal solution is 8%.

## 1.1 Motivation

According to today promotion in communication, IoT and 5G mobile network, demand for low-latency computation is increasing. MEC concept helps to reduce response time to MUs by providing cloud computing capabilities in the proximity of MUs. Moreover, by emerging new applications with higher required of computation like augmented reality and online gaming, workload offloading to MEC servers can reduce response time and increase QoE. Meanwhile, some

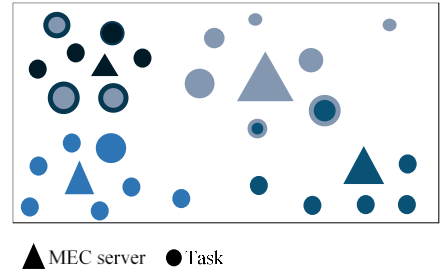

▲ MEC server   ● Task

**Fig. 2** – **Motivational Example:** The border and fill color of circles represents Host and Destination Servers, respectively.

MEC servers will get congested. So, migrating workloads can help using the free capacity of others.

We performed several experiments to observe the impact of migration on SLA violations. The setup contains 5 MEC servers with different configurations in a 1*0.5 km2 area. We spread 1000 requesting users uniformly in the area while each request can have different demands. Figure 2 illustrates an example where triangles and circles denote MEC servers and tasks, respectively. The variations in size correspond to computational capacity and demands. Border color of each circle represents the host MEC server. In the baseline where no migration happens, tasks must be processed in the host server. However, by enabling migration, each task can be processed on any MEC server. The fill color of circles represents the MEC server that is responsible for processing the task while workload migration is enabled.

Our experiments show that enabling migration reduces SLA violation by 30% on average. Because migration enables using the full potential of the network by preventing hotspots. These results motivate us to try to focus on presenting a right migration solution for scheduling and resource allocation of task among MEC servers to avoid hotspots.

## 1.2 Contribution
Our contribution is as follows:

- Formulate the task scheduling problem as a MINLP to minimize the SLA violation.
- Propose MESA as a greedy algorithm for finding a near-optimal solution in a much faster manner than the MINLP.
- On average, MESA has only 8% difference in term of SLA violation with the optimal solution.
- Study the effects of different parameters on SLA violation.

## 1.2 Organization
The rest of this paper is organized as follows. Section 2 presents a literature review of related works. We define the system model and formulation of the scheduling problem in Section 3. MESA is elaborated and analyzed in section 4. We present the simulation results in Section 5. Finally, conclude this paper in Section 6.

# 2 Related Work

The scheduling problem exists in many fields, even other than the computer science community, such as airline scheduling. Typically, this problem in computer science is to schedule a set of consumers, i.e., jobs, tasks, and threads to be executed on a set of servants, i.e., servers, CPU, and cores.

In computer science, many works [21]–[30] have done and used scheduling in different environments such as data centers, cloud, and embedded systems with different objectives such as makespan, energy consumption, power, and QoS. Some of them consider the scheduling of a job as a set of tasks, and some others consider multi-job scheduling.

Different environments have different criteria and challenges. In embedded systems, i.e., multi- and many-core systems, the servants and consumers are cores and threads, respectively, where the challenges are mostly the energy consumption and being real-time [31]. The computing part of a cloud data center can be formed from several embedded systems. The servants and consumers here could be workloads and servers. The challenge here is how to manage workloads to meet their requirements [32], [33]. Moreover, a cloud data center has some limitations, such as carbon footprint. Therefore, managers must have an eye on the energy consumption of the whole building as well [34].

Although there are some differences between the Cloud Data Center (CDC) and the Mobile Edge Cloud (MEC), both could be considered as networked computer systems. A significant difference is the type of workloads. In contrast to CDC, the workloads of MECs are mostly several jobs with short expected finish time less than a second while CDC workload may last some days. Another difference is topology and hierarchy of network. Each MEC server receives users' workloads and must respond to them, but in CDC the topology is master/slave.

Virtual machines (VM) frequently used in cloud data centers [22], [23], [28], [35]–[39]. They are emulations of physical machines (PM) that allows a PM to be considered as a set of machines with separated resources. First, a cloud data center must deploy VMs on PMs and then provide them tasks to execute them [33]. A technique called live VM migration is used frequently to consolidate VMs on a different and even smaller set of PMs [40]. The concept is similar to workload migration in MECs. The significant difference here is that we do not relocate the whole service presented by VM. We only migrate the workload, which is input data and a meta-data indicating the kind of service. For example, the data could be an image and the meta-data declaring that an edge detection filter should be executed on this data.

## 2.1 Scheduling in Mobile Edge Cloud

Here, the servants and consumers are MEC servers and workloads received from users on each MEC server. As we discussed in the introduction section, the research related to our work in this subsection can be divided into two categories: I) Workload offloading and II) Resource Management. In the following section, we will discuss related works of these groups.

**Workload offloading**

In this section, we focus on workload offloading decision on mobile user devices. It is a technique to solve the resource shortage problem of mobile devices. Workload offloading can be classified as static and dynamic methods. Static offloading means assign part of tasks to MEC servers at the start of application development statically [16]. However, static offloading is not suitable for mobile environment changes such as user mobility and network congestion. For example, authors in [14], [15] try to cope with energy consumption and try to offload the processing task to cloud servers dynamically and present an online algorithm that minimizes computation time. They take cloud service time into account and considers the general task graph for offloading.

The article [41] presents AppBooster, a mobile platform, which jointly leverages the quality adaptation, computation offloading, and parallel speedup to boost the overall performance.

Zhang et al. [42] proposed a combinational auction-based solution to select an in-range service provider for each mobile users to minimize the cost.

**Resource Management and Scheduling**

The focus of this group is on how to manage the system by proposing task scheduling and resource management techniques. In this step, the decision has been made about offloading of users' workloads, and they are of the givens of the scheduling problem [21], [43]–[45].

Lie et al. [20] proposed a scheduling for minimizing the average execution delay. They prioritize tasks by execution delay and energy consumption of tasks at the mobile device. In [46], authors deployed data-intensive edge service compositions on edge servers to combine several services with a logical integration under the driven data on edge servers. It can provide more powerful service functions, speed up system development, and meet user needs quickly.

Authors in [47] propose a power-efficient clustering scheme that minimizes the power consumption of MEC servers. Dinh et al. in [48] provide a solution to offload parts of a single user task to MEC servers. They try to minimize mobile device power consumption and the total task execution latency by selecting which part to execute on which server or on the mobile device. Chen et al. in [49] proposed a workload offloading policy at the mobile edge cloud. It decides that each mobile server connect to which in-range station to minimize the overall energy of servers.

Zhao et al. [50] consider a VM for each service in the regarding available services provided in 5G networks. Each task needs a VM. They proposed an algorithm to find a near-optimal placement of VM in the MEC servers to minimize the average response time. Likewise, in this work, we also consider that the network has some predefined services, and they are running on every MEC server.

Ceselli et al. [51] tackled the problem of designing a mobile edge cloud network while considering equipment, VMs, and users demands. First, they try to determine locations for radio stations and then decide each VM should place on which server to maximize the profit.

Authors in [52] introduce a time-slotted system and then, find CPU cycle needed for the task and find running time and schedule to maximize the profit of mobile service provider along optimizing power consumption. We used this system model in our work.

None of the above-mentioned works considers the workload migration concept. In this paper, we investigate the workload scheduling problem in a Multi-User-Multi-MEC manner with migration capability.

In our design, if a task enters MEC servers and finds it fully equipped, it will migrate to another MEC server. Each task has a specific expected response time. If it does not meet a deadline, an SLA violation will occur. The goal is to minimize SLA violations in each time slot according to task size, processing time, and the transmission time of task among MEC servers.

# 3 Problem Definition

We have several distributed heterogeneous MEC servers. These MEC servers have limited capacity, and each server receives tasks from their MUs. A controller server is responsible for gathering the information of tasks and capacity of the network, assigns tasks to suitable MEC server and resource management. We formulate this scheduling problem to a Multi-Integer Non-Linear Programming (MINLP) optimization problem with the objective of minimizing the SLA violations. Meanwhile, we consider workload migration among MEC servers.

Our MINLP optimization problem deals with a set of tasks $N = \{task_1, task_2, ..., task_{|N|}\}$ with different demands and a set of MEC server $M = \{MEC_1, MEC_2, ..., MEC_{|M|}\}$ with different configurations.

We used Million Instruction Per Second (MIPS) as the measure of processing power because for our model with heterogeneous servers. We need a single number to compare the systems as well as measuring the processing requirement of each workload, and the MIPS provides a reasonable estimate. There are applications like SimpleScalar [53] can give the number of instruction an application used to complete and also measure how many MIPS a server can execute.

Our assumptions are as follow:

1. There are a set of predefined services available on each MEC server.
2. Each task has a deadline for receiving the response.
3. Each task has a data and a meta-data indicating what service is required.
4. Only data and meta-data are required to execute a task on a MEC server and because the service is already running, the migration cost is all about transferring data and meta-data from a MEC server to another.
5. Time slot ($\tau$) is the clock of the system. Which means a decision for scheduling of incoming tasks is made at the beginning of each time slot.
6. At the beginning of each time slot, all scheduled tasks start computing in parallel.
7. Each task is assigned to only a single MEC server.
8. Each task should finish using one time slot.
9. A task can be delayed to next time slot if it did not miss the deadline already.
10. Computation capacity of each MEC server is limited.
11. The network topology is a connected component where all MEC servers are connected through a path.
12. The decision making performs for a group of distributed MEC servers.

## 3.1 Notation
We use the notation presented in Table 2.

## 3.2 Problem Statement
The given parameters of our problem are as follows:

- A set of tasks and their properties, e.g., arrival time, deadline, computational demands.
- A set of MEC servers and their properties, e.g. computational capacity.
- The properties of the network, e.g., the topology graph describing how MECs are connected.

Our goal is to minimize the SLA violations by determining each task to be computed on which MEC server while considering task migration among MEC servers.

## 3.3 Problem Formulation

First, we describe the total response time as follows:

$$TRT_i = t_i^{upload} + t^{decision} + t_{i,k}^{migration} + t_{i,k}^{process} + t_{i,k}^{response} \quad (1)$$

The meaning of each term is given in Table 2. The focus of this paper is to schedule latency-sensitive applications. Therefore, the values in the eq. 1 are expected to be in the order of milliseconds.

We ignored some delays in our model as they are negligible concerning maximum tolerable response time. However, tacking these delays into account will not hurt our model. We can take these delays into account as a constant since they do not have many variations and subtract the value from deadlines.

Now we formulate each time in $TRT_i$. The transmission time of task to associated MEC server is:

$$t_i^{upload} = \frac{s_i}{r_i} + \frac{d_{i,a[i]}}{v_c} \quad (2)$$

Where $s_i$ is input data size, $r_i$ is channel rate of $MU_i$, $d_{i,a[i]}$ is the distance between $MU_i$ and MEC server $a[i]$. Since the signal propagates through fiber optic media when transmitted among MEC servers, the second term converges to zero because of large denominator $v_c$ regarding the distance (speed of light = $3*10^8$ m/s).

Channel rate $r_i$ is calculated by Shannon–Hartley theorem [54]:

$$r_i = W \log_2(1 + \frac{h_i g_i}{I_i + \sigma_i^2}) \quad (3)$$

Where W is the total bandwidth between MUs and MEC server, $h_i$ is the transmission power of $MU_i$, $g_i$ is the channel gain of $MU_i$, $I_i$ is the interference of other MUs on $MU_i$, and $\sigma_i^2$ is the power of white noise of the environment.

After transmitting task to MEC server the time comes for the decision, whether process task on the current MEC server or migrate it to any other MEC server. The decision time does not affect solving the optimal solution as it is almost a constant and can be subtracted from the deadlines. Also, since we collect every task in the previous time slot and then, decide for next time slot, the decision-making process can be done slightly (a few milliseconds) before the start of next time slot.

If MEC server decides to migrate task to another server, the time of migration is calculated as follows:

**Table 2** Notations

| Symbol | Description |
|---|---|
| $N$ | Set of MUs tasks |
| $M$ | Set of MEC servers |
| $TRT_i$ | Total response time for task $i$ |
| $t_i^{upload}$ | Transmission time of sending task $i$ from MU to MEC servers on radio channel |
| $t^{decision}$ | Time of decision making for scheduling tasks |
| $t_{i,k}^{migration}$ | Transmission time task $i$ from MEC server $a[i]$ to MEC server $k$ |
| $t_{i,k}^{process}$ | Process time of task $i$ on MEC server $k$ |
| $t_{i,k}^{response}$ | Time of sending response of task $i$ to MU from MEC server $k$ to server $a[i]$ |
| $s_i$ | Size of input file of MU $i$ |
| $r_i$ | Channel rate of MU $i$ |
| $d_{i,a[i]}$ | Distance between MU$i$ to MEC server $a[i]$ |
| $a[i]$ | Indicate the host MEC server of task $i$ (int) |
| $v_c$ | Velocity of transmission media |
| $W$ | Total bandwidth among MUs and MEC server |
| $g_i$ | Channel gain of MU$i$ |
| $h_i$ | Transmission power of MU$i$ |
| $I_i$ | Interference of other MUs on MU$i$ |
| $\sigma_i^2$ | Power of white noise in environment |
| $R_{i,a[i],k}$ | Rate of link between MEC server $a[i]$ and server $k$ |
| $D_{a[i],k}$ | Distance between MEC server and server $k$ |
| $\alpha$ | Ratio of size of response file to input file |
| $s_i'$ | Size of file for response to MU$i$ |
| $X_{i,k}$ | To indicate whether task $i$ migrates from MEC server $a[i]$ to k or not (binary 0,1) |
| $L$ | A large number |
| $Y_i$ | Counter for deadline (SLA) violation |
| $t_i^{max}$ | Deadline time of task $a[i]$ |
| $f_{i,k}$ | Amount of MIPS assigned to task $i$ from MEC server $k$ (real) |
| $c_i$ | Million instructions needed for task $i$ (int) |
| $\tau$ | Time of each time slot (real) |
| $p_k$ | Computation power of MEC server $k$ in MIPS |

$$t_{i,a[i],k}^{migration} = \frac{s_i}{R_{i,a[i],k}} + \frac{D_{a[i],k}}{v_c} \quad (4)$$

Formula (4) is the summation of two terms. I) data transmission time of channel II) signal propagation time of the channel. Where $R_{i,a[i],k}$ is the rate of link between MEC server $a[i]$ and server $k$ and $D_{a[i],k}$ is the distance between MEC server $a[i]$ and server $k$. Note that $v_c$ is the speed of light which equals to $3*10^8 m/s$. Also, the network we consider is not larger than a city area. The widest distance between two points in three of vastest cities in the world, Sao Paulo, Tokyo, and Tehran, is not larger than 80km. Therefore, $D_{a[i],k}$ is at most 80km. So, the maximum value that the propagation delay can have is equal to 260us. Double this for a round trip is also a minimal value concerning maximum tolerable response time. So, we can omit propagation. Note to mention that even for these largest cities, we can use segmentation to not allowing a workload to migrate, for example, more than 10km.

After the task is placed on the destination, it will be processed. We set processing time such that it become done before the deadline and time-slot period by choosing a proper MIPS assigned to the task. According to Eq. (1) we can set $t_{i,k}^{process}$ as follow:

$$\forall i,k \quad t_{i,k}^{process} = \min(t_i^{\max}, \tau) - t^{decision} - t_i^{upload} - t_{i,k}^{migration} - t_{i,k}^{response} \quad (5)$$

Where $\tau$ is the size of each time slot. Therefore, the $\min(t_i^{\max}, \tau)$ term is the maximum acceptable $TRT_i$. This value maximizes the processing time while ensures no violation will happen in order to reduce the allocated processing power to maximize the number of tasks that can be executed in parallel on MEC server $k$. After the processing time is determined, we can find the required processing power allocated (MIPS) to the task $i$ on MEC server $k$ as follows:

$$f_{i,k} = \frac{c_i}{t_{i,k}^{process}} \quad (6)$$

Where $c_i$ is million instruction demanded by task $i$.

Afterward, the response will be sent back to MU. The response file size ($s_i'$) depends on input file size and the application. It is calculated as follows:

$$s_i' = \alpha s_i \quad (7)$$

Where alpha is a ratio related to the application.

Now, this file has to return to associated MU along the path that task was sent by MU. Response time consists of two terms. Transmission time to MEC server $a[i]$ and time of receiving a response by MU.

$$t_{i,a[i],k}^{response} = \frac{s_i'}{R_{i,a[i],k}} + \frac{D_{a[i],k}}{v_c} + \frac{s_i'}{r_i} \quad (8)$$

The last term is the time of delivering the result from the host MEC server to MU where $r_i$ is channel rate of $MU_i$. We can omit propagation time (second term) because it is converging to zero.

We describe a binary variable $X_{i,k}$ to show that task $i$ is executing on MEC server $k$ as follows:

$$X_{i,k} = \begin{cases} 1 & \text{if task } i \text{ assigned to server } k \\ 0 & \text{otherwise} \end{cases} \quad (9)$$

Note that $X_{i,a[i]} = 1$ means the task is executing on the host MEC server.

For finding an optimal solution, we can rewrite $TRT_i$ as follow:

$$\forall i \in N$$

$$TRT_i = t^{decision} + t_i^{upload} + \sum_{k=1}^{|M|} X_{i,a[i],k} \left( t_{i,k}^{migration} + t_{i,k}^{process} + t_{i,k}^{response} \right)$$

$$+ L(1 - \sum_{k=1}^{|M|} X_{i,k}) \quad (10)$$

Where L is a large number that is bigger than the maximum deadline causing an SLA violation when a task is not assigned to any MEC server.

We describe $Y_i$, a binary variable indicating task $i$ violation as follows:

$$Y_i = \begin{cases} 1 & TRT_i > t_i^{max} \\ 0 & \text{otherwise} \end{cases} \quad (11)$$

Now, we formulate our mixed-integer non-linear optimization problem where $X_{i,k}$ and $f_{i,k}$ are constraint variables:

Minimize $\sum_{i=1}^{|N|} Y_i \quad (12)$

s.t. $X_{i,k} \in \{0,1\} \quad (13)$

$Y_i \in \{0,1\} \quad (14)$

$$\forall i,k \quad t_{i,k}^{process} = \min(t_i^{\max}, \tau) - t^{decision} - t_i^{upload} - \sum_{k=1}^{|M|} X_{i,k}(t_{i,k}^{migration} + t_{i,k}^{response}) \quad (15)$$

$$\forall i,k \quad f_{i,k} = \frac{c_i}{t_{i,k}^{process}} \quad (16)$$

$$\forall k \quad \sum_{i=1}^{|N|} X_{i,k} \times f_{i,k} \leq p_k \quad (17)$$

$$\forall i \quad \sum_{k=1}^{|M|} X_{i,k} \leq 1 \quad (18)$$

Constraint (13) determines the MEC server that should execute task $i$. Constraint (14) shows that whether an SLA violation happens or not.

As we explained earlier, $t_{i,k}^{process}$ is calculated based on the remaining time of deadline and considered by constraint (15). Constraint (16) demonstrates the amount of MIPS that each task should be given to guarantee that all scheduled tasks meet their deadlines. Constraint (17) ensures that processing power that allocated to scheduled tasks on a MEC server should not be larger than its capacity ($p_k$) during of time slot τ. Constraint (18) shows that no task can be assigned to more than one MEC server. Finally, minimizing SLA violations ($Y_i$) is done by this optimization problem (12).

# 4 Algorithm

As we mentioned before, the mixed-integer non-linear problem has immense time complexity, and finding the optimal solution on a large scale is not feasible. Also, according to Table 1, the latency of applications needs a response in order of milliseconds while solving an optimization problem at the beginning of each slot is impossible. So, in this section, we introduce a greedy algorithm called MESA, for workload scheduling and investigate its complexity. MESA runs much faster than solving the optimization problem and provides near-optimal solutions.

## 4.1 Algorithm Definition

The inputs of MESA are all information about the task, mobile users, the network, and MEC servers. MESA is an algorithm that ensures every constraint described in the previous section. It first prioritizes tasks by considering two factors: I) The number of million instruction demanded by each task and II) maximum acceptable response time of each task ($t_i^{max}$). Hence, we consider a new parameter $\phi_i$ as follows for each task (line 2):

$$\phi_i = \frac{c_i}{\min(t_i^{max}, \tau)} \quad (19)$$

It sorts task according to $\phi_i$ in ascending order (line 3), which means all tasks are sorted regarding an approximate of their demanded MIPS. This policy gives tasks with lower demand higher priority to ensure the maximum number of tasks that are executed to minimize the SLA violations percentage.

MESA peek tasks one by one for the list of tasks $N$ (line 4). Then, select a MEC server to execute the task. For doing that, it first sorts MEC servers regarding their distance to the host MEC server (line 5). Note that the host MEC server always is on the top of the list as it has no distance to itself. Therefore, for each task, it first investigates the feasibility of processing the task on the host MEC server. If there is enough processing power, the task will be assigned to the host MEC server. Otherwise, it considers other MEC servers (lines 4-13). The process of selecting another MEC server is such that it gives higher priority to nearest MEC servers. This will minimize the migrating time as well as network usage.

**MESA Algorithm**

**Inputs:** $M, N, D, R, p_k, c_i, s_i, s'_i, h_i, r_i, g_i, \alpha, I_i, v_c, \tau, t_i^{max}$

**Outputs:** $\{X_{i,k}\}, \{f_{i,k}\}, Y_i$

1. $Y_i = 0$ //as a counter of SLA violation
2. Calculate $\phi_i = \frac{c_i}{\min(t_i^{max}, \tau)}$ for each task
3. Sort $N$ in ascending order with respect to the value of $\phi_i$
4. for $i$ from 1 to $|N|$:
5.     $Servers$ = sort $M$ in ascending order w.r.t. the distance
6.     for $k$ in 1 to $|Servers|$ do:
7.         $t_i^{process} = \min(\tau_i^{max}, \tau) - t^{decision} - t_i^{upload} - t_{i,k}^{migration} - t_{i,k}^{response}$
8.         $f_{i,k} = c_i / t_i^{process}$
9.         if $p_k > f_{i,k}$ :
10.             $X_{i,k} = 1$ //assign request$_i$ to MEC server $k$
11.             $p_k = p_k - f_{i,k}$
12.             Break
13.     End for
14.     if $\sum_{k=1}^{|M|} X_{i,k} = 0$ :
15.         $Y_i$++
16.     End if
17. End for
18. Calculate TRTs based on the decision (X)
19. for $i$=1 to $|N|$ do:
20.     if $TRT_i > t_i^{max}$:
21.         $Y_i$++
22.     End if
23. End for
24. Return: $\{X_{i,k}\}, \{f_{i,k}\}, Y_i$

If MESA fails to find a MEC server with enough processing capacity for a task, a violation happens (line 14-16). Afterward, it calculates TRTs based on the decisions according to Eq. 10. (line 18). Finally, it counts the number of tasks that violate the SLA (line 19-23) and returns the schedule (line 24).

## 4.2 Complexity of Algorithm

The complexity of filling Arrays $R$ and $D$ is $O(|M|^2+|N|+|M|)$, where $|M|$ is the number of MEC servers and $|N|$ is the number of tasks. The complexity of calculating parameter $\phi_i$ for each is $O(|N|)$ (line 2). Afterward, MESA sorted tasks with Merge Sort algorithm with the complexity of $O(|N| \log|N|)$ (line 3).

In the main loop of the algorithm (line 4-17), another Merge Sort is performed for sorting MEC servers with the complexity of $O(|M| \log |M|)$ (line 5). There is another *for* loop (line 14-16) in the main loop for checking an SLA violations occurrence with the complexity of $O(|M|)$. So, the main loop complexity is $O(|N||M| \log|M| + |N||M|)$. Finally counting SLA violations is a *for* loop with $O(|N|)$ (lines 19-23). Thus, the total complexity of the algorithm is

$O(|M|^2+3|N|+|M|+|N| \log|N|+|N||M| \log|M| + |N||M|)$.

Because of $|M|^2 << |N|$, it can be concluded that the total complexity of MESA is $O(|N||M| \log|M| +|N| \log|N|)$.

# 5. Evaluation

In this section, we evaluate the performance of MESA. For a fair comparison, we searched for works with our objective while considering workload migration, and we found none. Therefore, we compared MESA with the optimal solution and a randomized algorithm. The optimal solution is gained by solving the model using optimization solver in GAMS software, and the randomized one is a solution that randomly decides migrations and resource allocation. We run each algorithm 20 times and report the average as the answer. First, the architecture and configuration of the simulation are presented. Then, experiments and results are investigated.

## 5.1 Experiments Setup

As mentioned in section 1, we use MEC architecture for this simulation, which means it includes MUs and MEC servers and the central data center. In this problem workload scheduling decision is made by an aggregation node (shown in Fig. 1) that includes a controller server. In each time slot, MEC servers send meta-data of their tasks to the controller server. This meta-data includes task file size ($S_i$), million instruction of task ($c_i$), entrance place of the task ($a[i]$), and deadline of the task ($t_i^{max}$). Since this information is minimal.

So, transmission time to the controller is negligible. Finally, controller schedule tasks for all MEC servers.

For solving the optimization problem, we used the GAMS tool, which is a comprehensive system for mathematical modeling problems primarily linear, non-linear, and mixed. We used BONMIN solver for solving the MINLP problem. Simulating of network and MESA is done by C++ programming.

All of the experiments are done by a system with 16 Gigabyte RAM, Intel Core i7-6650 processor with frequency 2.2 GHz. Datasets and information that are used in the experiments are shown in Table 1. For the size of input tasks, we use the normal distribution. We set mean and variance according to [48], [55] with a mean of 23000 million instructions and variance of 3500. According to Table 1 and latency-sensitive application, the default time slot is set to 2 seconds, but we will also consider other time lengths in a separate experiment.

**Table 3** Topologies.

| Graph | Links | Nodes |
|---|---|---|
| Renam | 4 | 5 |
| CESNET | 9 | 10 |
| SpiraLight | 16 | 15 |

Network graphs and topologies are taken from The Internet Topology Zoo dataset [56] which are real network topologies and gathered by the University of Adelaide. These graphs are presented in Table 3, which includes network links and nodes. Nodes are MEC servers and are heterogeneous, which means MIPS for each of them are different. We use values in Mega MIPS that comes from processing frequency of MEC server in [48]. Radio channels have transmitting power 1.5 watts, and white noise of the environment is -60 dB [57]. The topology of experiments is SpiraLight with 15 nodes and 16 links.

## 5.2 Results

We designed different experiments to evaluate the performance of MESA.

**Processing Power**

In our first set of experiments, we investigate the effect of increasing processing power by increasing the number of MEC servers for a set of fixed tasks. The results are shown in Fig. 3 with an increase in the number of MEC servers, deadline violation decreases. It shows the ability of our model that can use every MEC servers processing power by enabling migration. As it is shown, MESA has only 4% difference with optimal. In this experiment, the maximum difference between MESA and optimal is 11%, where the number of MEC servers is 5. We can perform this type of experiments to suggest a suitable number of MEC server in a fixed area for an approximation of workload in order to provide a good trade-off between violation and cost of equipment of MEC servers.

We then designed another set of experiments that increase in the number of MEC servers does not change the total processing power provided by MEC servers. Figure 4 shows the SLA violation of 3000 tasks that are distributed in a 2.5*5km$^2$ area, while the locations of MUs are chosen randomly. We have fixed processing power, but it is divided between MEC servers. According to results with increasing the number of MEC servers, deadline violation percent increases slowly. The increase happens because the fixed processing power among more MEC servers become fined-grained, and task with high processing demands will not be served. The remarkable point of this experiment is the slowly increasing of violations, which means that it is better to use more but cheaper MEC servers instead of fewer expensive ones. Since with vertical scaling of processors, the cost will increase exponentially, using a higher number of cheap servers is more economical.

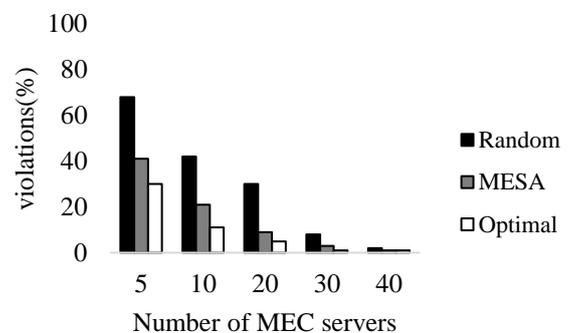

**Fig. 3** Deadline violation for 3000 tasks (increasing processing power)

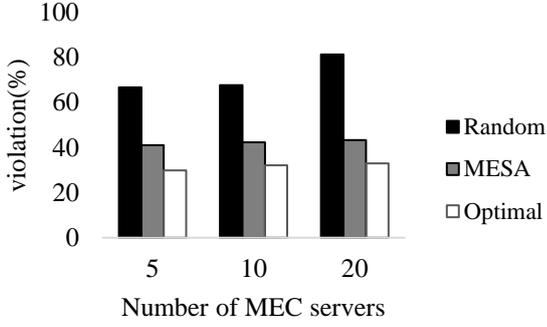

**Fig.4** Deadline violation for 3000 tasks (fixed processing power)

### Number of Tasks

Figure 5 shows the results of experiments with an increasing number of tasks with fixed processing power. Here we consider a fixed number of MEC servers (15). The area is 5*2.5 km$^2$. Increase in the number of tasks causes more deadline violation. Results show that the performance of MESA and difference with the optimal solution is on average 7%. As the number of tasks increases, MEC servers accept and process tasks until they are saturated. After this moment, every task enters the MEC server will be rejected, and deadline (SLA) violation happens. With this simulation, we can estimate the number of tasks that can receive a response in each time slot. In other words, we can define the capacity of each group of MEC servers.

### Runtime

We demonstrate the run time of MESA comparing with finding the optimal solution in Table 4. As it shows, the run time of the optimal solution is much more than MESA. Because it is a MINLP optimization problem and it is not scalable. For example, the run time of MESA and optimal solution to schedule 1500 tasks are respectively 8.9 and 850000 milliseconds, which is about 14 minutes. According to time slots duration and the maximum deadline for latency-sensitive applications solving optimization problem is not acceptable.

### Time Slot

The time slot period is a parameter directly related to the distribution of workloads. When the services are predefined, and the area is fixed, it is expectable that the workload distribution is predictable based on history. So, one can use our model to find the best time slot for a period of time.

As mentioned earlier, we test a range of different time slots for our experimental set. Figure 6 shows the effect of different time slots on SLA violations. Time slot range that we used is between 0.5 and 4 seconds with 0.5-second steps. Starting from 0.5 seconds, with increasing duration of slots SLA violation decreases because there will be more room for larger tasks, but it will reach a minimum spot that violation is increasing afterward. It is because of the violation of smaller tasks as they must wait until the start of the next time slot where they might miss the deadline even before being scheduled. The minimum spot is related to the distribution of the workload. The difference between optimal solution and MESA is 8% on average in these experiments.

**Table 4** Runtime of MESA and optimal solution with 15 MEC server

| Number of Tasks | Optimal (ms) | MESA (ms) |
|---|---|---|
| 500 | 270000 | 8.5 |
| 1000 | 410000 | 8.8 |
| 1500 | 850000 | 8.9 |
| 2000 | 1630000 | 8.9 |
| 2500 | 4020000 | 9.2 |

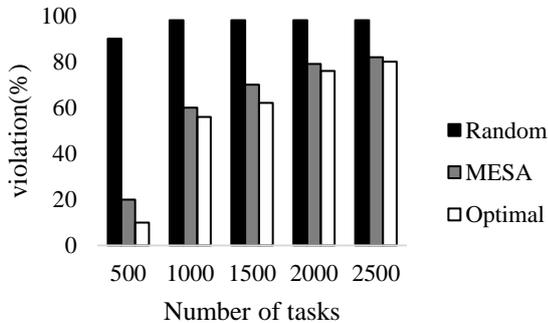

**Fig. 5** Deadline violation to the number of tasks with 15 MEC servers

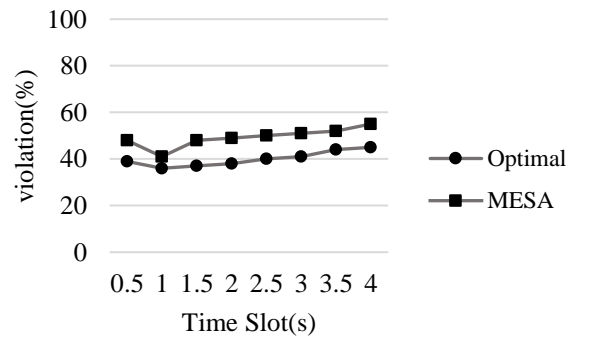

**Fig. 6** Deadline violation to time slot size with 15 MEC servers

**Distribution of Tasks**

In the next experiment, we evaluate the effect of mobile user distribution over a fixed area. We consider a 1*0.5km2 area, 5 MEC servers, and 1000 tasks. Figure 7(a) shows a uniform distribution of task and Fig. 7(b) demonstrate non-uniform distribution, which density of tasks over two MEC server is much more. SLA Violation percentage for the uniform and non-uniform distribution of tasks are shown in Fig. 8. A case of without migration is also considered here, which means if there is no capacity on the MEC server for a task deadline violation will happen. In non-uniform distribution, the difference between MESA and 'without migration' is much higher, because we can migrate tasks from hotspot MEC servers to others to use their processing power. In general, deadline violation for uniform distribution of tasks is lower than non-uniform one, because of the lower transmission time of the task. In non-uniform case, tasks will be sent to long-distance MEC servers that may cause much time and deadline violation. We conclude that the migration of tasks and use other's capacity will reduce deadline violation. On average, the difference between optimal and MESA is 8.6%. Finally, the results show good robustness of our proposed model and MESA regarding the hotspots.

**Different Topologies**

Figure 9 illustrates the effect of different topologies on SLA violation. These experiments consider 1000 tasks, scheduled on three different topologies. As it is shown, the violation is lower in the topologies with better network coverage. These

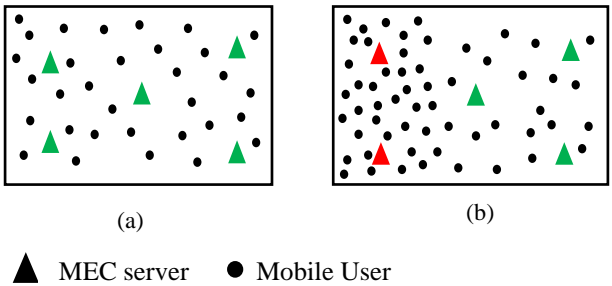

▲ MEC server   ● Mobile User

**Fig. 7** Distribution of tasks uniform (a) and non-uniform distribution with two hotspots (b)

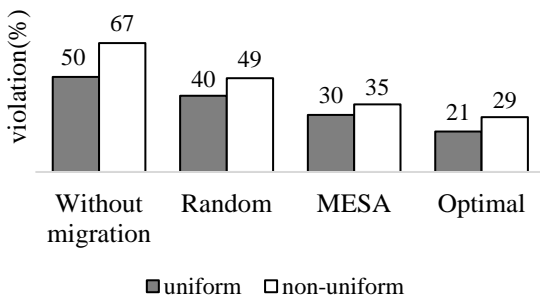

**Fig. 8** Deadline violations with 5 MEC servers with task distribution

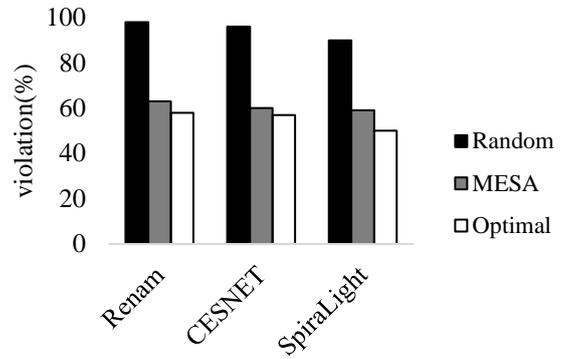

**Fig. 9** Violation percent based on different topologies

experiments show the robustness of MESA regarding the topology as it performs near-optimal in different networks.

## 6. Conclusion

According to the next mobile Internet generation 5G and its properties like low latency, Mobile Edge Cloud is an opportunity that computes mobile user's task on the edge of the network and reduces latency. Furthermore, new applications consume more energy that makes workload offloading indispensable to save battery. In this paper, we consider a MEC network with several MEC servers besides mobile users that receives workload from users and computes them. Also, each user has a specific SLA and maximum waiting time to receive its response. Our goal is to minimize SLA violation and increase users' satisfaction by scheduling tasks on MEC servers with workload migrating consideration. We modeled this problem as a mixed-integer non-linear programming model for obtaining the optimal solution. Since solving this kind of problems are time-consuming (in order of minutes) even on a small-scale network, we introduce a greedy algorithm, MESA, with much lower complexity (in order of milliseconds). MESA schedules the tasks and manages the resources of MEC servers and makes decisions based on properties of tasks and servers to meet maximum acceptable response time of tasks. Solutions of MESA are 8% and 11% far from the optimal on average and in the worst case, respectively.

Our future work has three directions: I) considering the problem as a multi-objective problem, II) consider multiple task distributions, and III) considering more details. For the first, we planned to redesign the model to provide a good trade-off between several objectives such as SLA violation, network usage, and energy consumption. In this work, we consider all the workloads came from the same distribution. Our future plane for the second direction is to consider workloads from different classes of distributions. For the third, we planned to make the model more complex by considering more details, such as giving the tasks priority level.